\documentclass[twocolumn,english,aps,superscriptaddress]{revtex4}
\usepackage[T1]{fontenc}
\usepackage[latin9]{inputenc}
\usepackage{graphicx}
\usepackage{esint}
\usepackage{color}
\usepackage{babel}

\makeatletter

\providecommand{\tabularnewline}{\\}

\@ifundefined{textcolor}{}
{%
 \definecolor{BLACK}{gray}{0}
 \definecolor{WHITE}{gray}{1}
 \definecolor{RED}{rgb}{1,0,0}
 \definecolor{GREEN}{rgb}{0,1,0}
 \definecolor{BLUE}{rgb}{0,0,1}
 \definecolor{CYAN}{cmyk}{1,0,0,0}
 \definecolor{MAGENTA}{cmyk}{0,1,0,0}
 \definecolor{YELLOW}{cmyk}{0,0,1,0}
}

\makeatother

\begin{document}

\title{MC-DEM: a novel simulation scheme for modeling dense granular media}

\author{Nicolas Brodu}
\affiliation{INRIA, 200 avenue de la Vieille Tour, 33405 Talence, France.}
\affiliation{Dept. of Physics \& Center for Nonlinear and Complex Systems, Duke University,  Box 90305, Durham, NC 27708-0305, USA}
\author{Joshua A. Dijksman}
\affiliation{Department of Physical Chemistry and Colloid Science, Wageningen University, PO Box 8038, 6700EK Wageningen, The Netherlands.}
\affiliation{Dept. of Physics \& Center for Nonlinear and Complex Systems, Duke University,  Box 90305, Durham, NC 27708-0305, USA}
\author{Robert P. Behringer}
\affiliation{Dept. of Physics \& Center for Nonlinear and Complex Systems, Duke University,  Box 90305, Durham, NC 27708-0305, USA}

\begin{abstract}
This article presents a new force model for performing quantitative
simulations of dense granular materials. Interactions between multiple
contacts (MC) on the same grain are explicitly taken into account. Our
readily applicable method retains all the advantages of discrete element method (DEM) simulations
and does not require the use of costly finite element methods. The new model
closely reproduces our recent experimental measurements \cite{3DXP},
including contact force distributions in full 3D, at all compression
levels up to the experimental maximum limit of 13\%. Comparisons with 
traditional non-deformable spheres approach are provided, as well
as with alternative models for interactions between multiple contacts.
The success of our model compared to these alternatives demonstrates
that interactions between multiple contacts on each grain must be
included for dense granular packings.
\end{abstract}
\maketitle

Dense particulate media such sand, emulsions and colloids are
ubiquitous in nature and in industry. However, understanding their
very rich mechanical behavior has been notoriously
difficult. Numerical simulations are an essential tool to access the
microscopic and macroscopic behavior of these systems. In principle,
the application of solid mechanics and Newton's laws of motion to
every grain in a packing should recover that packing
macroscopic behavior. These simulations are typically referred to as Discrete Element Methods (DEM)~\cite{DEM_recent} and tremendous progress has been made since the classic work of Cundall and Strack~\cite{DEM_ori}. 
However, getting quantitative agreement between experimental results and DEM
simulations is often a challenge. This is partly due to the
lack of microscopic structure and force data in experiments on dense
particulate media; usually only boundary stresses are available.
In the last decade however, much progress has been made in
obtaining microstructural data in two and three dimensional model
experiments in emulsions \cite{Brujicx2,droplets} and granular materials
\cite{GEM,elastic3D,triax}. We have recently experimentally
measured all grain-scale properties for a 3D granular system,
including inter-particle contact forces, as the system
was subject to controlled strain \cite{3DXP}. These experimental
approaches provide a testing ground for DEM models. We show that
conventional DEM methods need modification to give a good quantitative
match to our recent 3D experimental data.  

We propose a novel method dubbed multi-contact DEM (MC-DEM), 
that vastly improves the predictive power of DEM methods, while
retaining their conceptual simplicity. The
essential ingredient is that our method computes an overall grain
shape deformation induced by particle contacts. This allows for two
key effects that are not taken into account in traditional DEM:
\textit{(i)} shape deformations induce the formation of new contacts;
\textit{(ii)} every contact force is affected by other contact forces
in its vicinity. We show that our method is in very good agreement
with our recent 3D experimental results. Importantly, our method does not
involve finite element method (FEM) calculations.  Particle shape
deformations are computed with analytic linear elasticity
calculations. More generally, the same method should be applicable to all particulate media where a relation between shape and deformation is readily available, such as
for emulsions or foams~\cite{foams}.

\begin{figure}
\includegraphics[width=1\columnwidth]{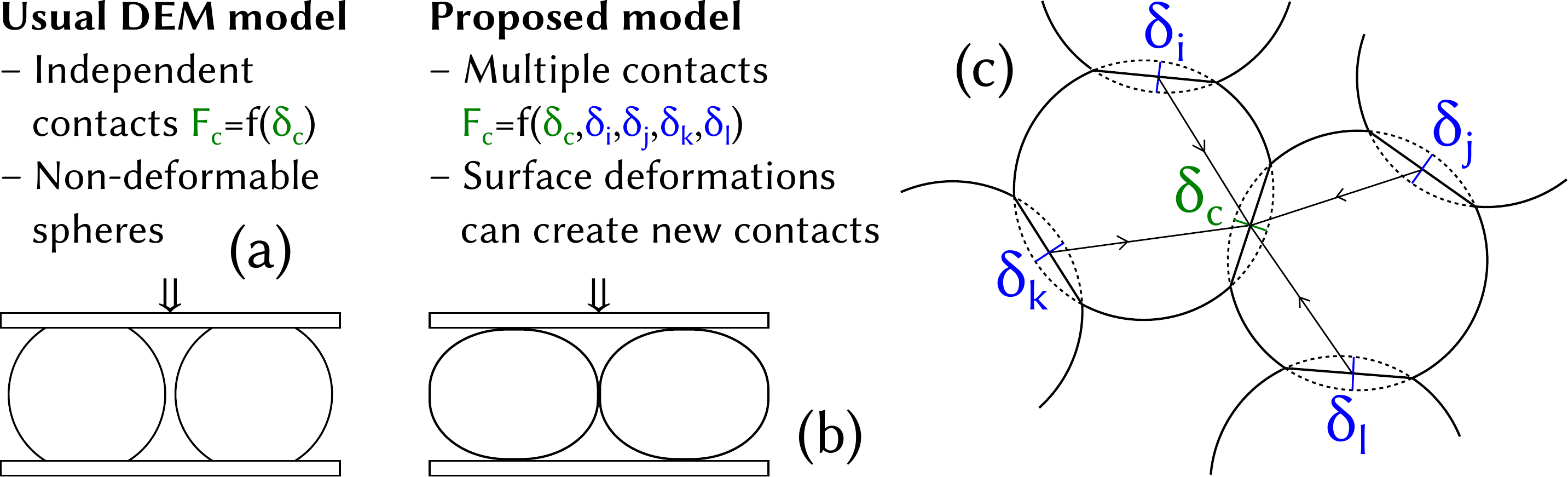}

\caption{\label{fig:multiple_contacts} When squeezing a packing of particles, particles deform and new contacts are formed. Traditional DEM (a) does not take this into account. Our multiple contacts model (b) captures these by correcting contact deformation $\delta_c$ using linearly additive corrections from other contacts $\delta_{k\rightarrow c}$, making (c) the local force depend on all particle contacts $F_c = f(\delta_c,\delta_i, \delta_j,\delta_k,\delta_l)$.}
\end{figure}

\section{Background}
The conventional DEM approach \cite{DEM_recent} assumes that all
contact forces are binary, i.e. independent of forces acting elsewhere
on a particle.  This is not rigorously true, since the force at one
contact induces strains throughout the particle, including at other
contacts. In turn, any modification of a force at a second contact due
to a force elsewhere on a particle propagates throughout the system
due to Newton's third law. The binary assumption avoids computational
complexity, and makes conventional DEM feasible. The main ingredient
in our new DEM method is to add interactions between multiple contacts
on the same grain, as sketched in Fig. \ref{fig:multiple_contacts},
but without incurring long range effects.

Typically, the normal component of a DEM force law is described in
terms of the `overlap', $\delta$ of grains, and the resulting force is
proportional to $\delta ^{3/2}$ for Hertz' law, or to $\delta^1$ for
linear springs.  Depending on the model, tangential forces are added
to account for surface friction, plasticity, cohesion, etc. Without
correlation between contacts, this method is similar in spirit to a
mean field approach and provides only a first order description of the
granular assembly properties. Hence, a more accurate effective
multi-contact force model must correct for this, by including the
coupling of forces at multiple contacts on a given particle, which is
what we do in our model. Of course, for granular gases~\cite{granugaz}, or
loose granular flows~\cite{granuflows}, binary collisions are a good
approximation. The situation is different in dense granular packings
near or above jamming. Here, force propagation through long-lasting multiple
contacts per particle are the norm, and clearly highly relevant for
dynamics~\cite{dynamics,nlsoundimpact}. 

\section{Introducing Multiple Interactions}
\begin{figure}
\includegraphics[width=0.75\columnwidth]{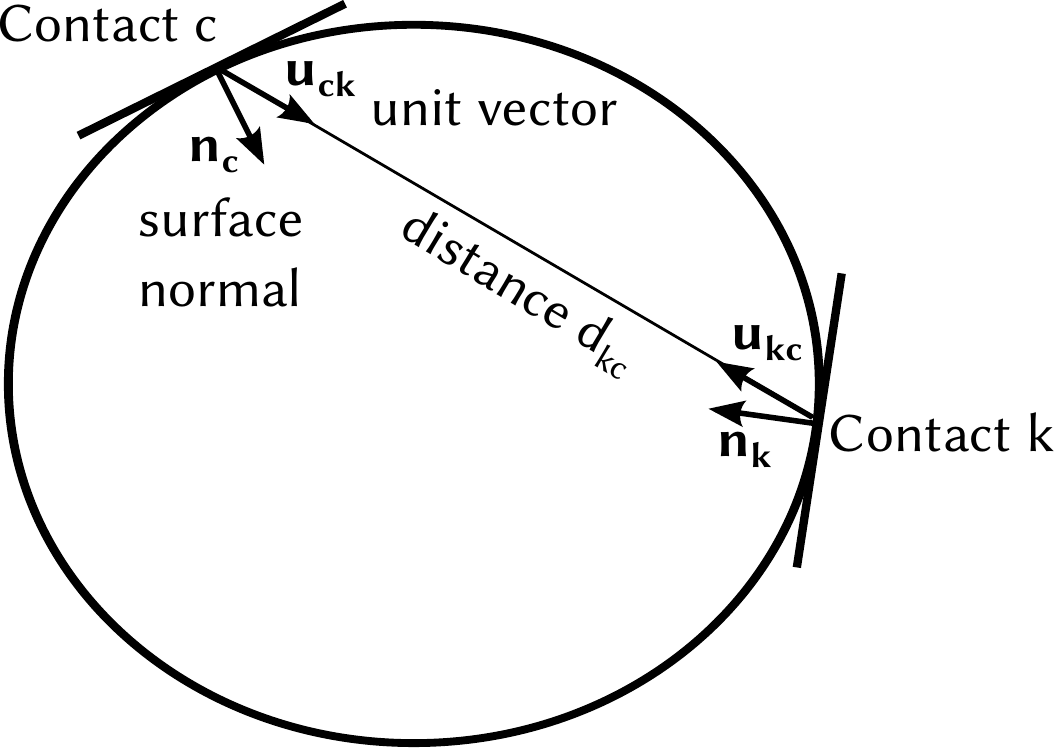}
\caption{\label{kc_geometry}Influence of one contact onto another. Contacts
are not restricted to the surface of a sphere, their position is consistent
with the grain deformations.}
\end{figure}

Interaction schemes for multiple contacts have been proposed
\cite{multicontacts}, but not in the context of DEM. The idea is to
model the mutual influence of contacts. This is done by
using information on deformations induced by one contact force on the
other contacts acting on the grain (Fig.~\ref{fig:multiple_contacts}c). The displacement fields $\delta_{k\rightarrow c}$ in the
normal direction, induced by the deformations at other contacts
$\delta_{k}$, are then added to the particle deformation at the local
contact $\delta_{c}$~\cite{multicontacts}, before applying Hertz' law
$F\propto\left(\delta_{c}+\sum_{k}\delta_{k\rightarrow
  c}\right)^{3/2}$. At this point, various choices may be made for the
form of that displacement field $\delta_{k\rightarrow c}$. A first
choice is to use the solution for a point force on a sphere
\cite{point_sphere_contact}, which we consider below. A second choice
is to consider the stresses induced by an extended surface contact
between two spheres \cite{sphere_sphere_contact}, which is
approximated in a simpler form in \cite{multicontacts}.
Unfortunately, while these models may work well for spheres in
isolation (i.e. dilute flows, where the contact models matters less),
there is no theoretical justification for choosing a solution based on
a spherical boundary condition for dense packings, where that boundary
may be a very complicated set of free surfaces between contacts. In
this work, we therefore simply consider the solution for a point force
on an infinite half-space, arguing that the dense packing more closely
resembles this situation than that of an isolated sphere. In the quantitative
comparisons with experimental data below, we find that our effective infinite half
space approach works better than the other particle-level perspectives.
Using the notation of Fig.~\ref{kc_geometry}, we express the point force on the elastic half-space solution from \cite[5.4.4]{solid_mechanics} in a vectorial form as:
\begin{eqnarray}
\delta_{k\rightarrow c} & = & -\gamma\frac{(1+\nu)F_{k}}{2\pi Ed_{kc}}\left\{ \vphantom{{\frac{\left(\mathbf{n}_{k}+\mathbf{u}_{kc}\right)\cdot\mathbf{n}_{c}}{1+\mathbf{n}_{k}\cdot\mathbf{u}_{kc}}}}\left(\mathbf{n}_{k}\cdot\mathbf{u}_{kc}\right)\left(\mathbf{n}_{c}\cdot\mathbf{u}_{kc}\right)\right.\label{eq:deltakc}\\
 &  & \left.+(3-4\nu)\mathbf{n}_{k}\cdot\mathbf{n}_{c}-\left(1-2\nu\right)\frac{\left(\mathbf{n}_{k}+\mathbf{u}_{kc}\right)\cdot\mathbf{n}_{c}}{1+\mathbf{n}_{k}\cdot\mathbf{u}_{kc}}\right\} \nonumber 
\end{eqnarray}

\noindent
with $F_{k}$ the force at contact $k$, $E$ the Young modulus of
the material, $\nu$ its Poisson ratio. If the contacts $k$ and $c$
are restricted to be exactly on the surface of a non-deformable sphere,
and for $\nu=0.5$ (and only that case) the half-space solution is
the same as in \cite{multicontacts}. In practice, it is simpler
to implement the linear elasticity solution in a simulation: the cross-contact
influences are computed with Eq.~\ref{eq:deltakc}, considering the
current positions of contact points as in Fig.~\ref{kc_geometry}, without having to resort to a
spherical approximation. However, in a dense packing, the grains do
not form an infinite continuous half-space, but rather a very complicated
and ever-changing boundary with pores between grains. For non-compressible
materials (Poisson ratio $\nu=0.5$) the solution for an infinite
full-space \cite[5.4.3]{solid_mechanics} only induces a change in
prefactor, $\gamma=0.5$ instead of $\gamma=1$ in Eq.~\ref{eq:deltakc}.
We therefore let $\gamma$ be an adjustable parameter in order to
account empirically for the geometry. 

\section{Models}
We consider the following models:

\textendash{}~\emph{Independent contacts}: The reference model with
non-deformable overlapping spheres \cite{DEM_ori} and independent
contacts with Hertz' law.

\textendash{}~\emph{Point force on sphere}: Bondareva's solution
\cite{point_sphere_contact} for the deformations induced by a point
force on a sphere. The radial component of Eqs 7 and 8 in \cite{point_sphere_contact}
is used for $\delta_{k\rightarrow c}$.

\textendash{}~\emph{Sphere to sphere}: The approximate solution by
Gonzalez and Cuitiño \cite{multicontacts} of a sphere-to-sphere contact
\cite{sphere_sphere_contact}; $\delta_{k\rightarrow c}$ is then
set to Eq.~3 in \cite{multicontacts}.

\textendash{}~\emph{Elastic Half Space}: The above linear elasticity
solution, Eq.~\ref{eq:deltakc}, for the deformations induced by a
point force on an infinite half-space, with $\gamma=1$. As noted
above, when $\nu=0.5$, the only difference with the sphere to sphere
solution is that contact points are applied at the location of the
deformed grain surface, rather than being projected on a sphere.

\textendash{}~\emph{MC-DEM}: The full Eq.~\ref{eq:deltakc},
adjusting $\gamma$ so as to best match experimental measurements.

At each simulated time step, the total displacements
$\sum_{k}\delta_{k\rightarrow c}$ are computed not only at the contact
points, but also at the surface of each grain in the directions of the
closest neighbors (Fig.~\ref{fig:multiple_contacts}). When these
surface deformations are large enough that the particles would
overlap, a new contact is created. That contact initially produces a
zero force when the deformed surfaces barely touch. Hence there is a null
cross-contact influence according to Eq.~\ref{eq:deltakc}, so the
creation of new contacts is a continuous process. This method
naturally handles the automatic creation of new contacts due to
particle deformations, which the basic DEM model is unable to
achieve. With the same Young modulus for each material in contact, the
amount of deformation spreads equally in both particles
\cite{LnL}. Therefore, the contact position is geometrically defined
as the average between the surface positions if they were overlapping.
Unlike the standard DEM, where spheres are not deformable, the radius
now effectively changes per contact. This yields geometric torques
$f_{t}\times r$ due to non-sphericity, that are handled at no
additional cost. In the standard DEM, only $f_{t}$ varies with
friction, while now $r$ changes as well. However, recomputing the
inertia matrix of each grain using the surface deformations would be
very costly, so we keep the inertia of a sphere. With many contacts
spread around the grain, we assume that the situtation is isotropic
enough that the sphere inertia is a reasonable approximation. 
Thus, only the
$\delta_{k\rightarrow c}$ computations themselves significantly add to
the simulation cost as $O\left(ZN\right)$, where $Z$ is the number
of contacts per grain and $N$ the number of neighbors per grain.

\section{Experiment}

Our reference experiment consists of 514 hydrogel grains that are
uniaxially compressed 20 times consecutively \cite{3DXP}. We have
measured the system at each compression level, from strains of 0 to
$13.4$\%.  Each compression/unloading cycle comprises 50 full 3D scans
where the top plate touches the grains, with 10 additional scans where
the top plate is above the packing. We discard the first 5 compression
cycles to let the system reach a reproducible configuration from cycle
to cycle. We explicitly separate loading and unloading phases as some
hysteresis is observed, which the simulations also reproduce. The
full 3D force vectors are available experimentally at each contact
\cite{3DXP}. In addition, we also measure the force exerted on the
top plate at each scan. These statistics, averaged over all similar
loading/unloading phases, form the ``ground truth'' that the DEM
models must reproduce as accurately as possible.

The experimental particles are not completely spherical in their
uncompressed state. In the simulations, we replace them by spheres
with the same volume and same initial center of mass, and let the packing
relax to a nearby state. This process yields a slight inflation of the
simulated packing, which generates a non-zero force on the top
plate in the least compressed samples, unlike the experiment.  This
effect is negligible in the most compressed states. We use our best
estimate of the hydrogel Young modulus $E=23.3$~kPa, and a constant
coefficient of restitution \cite{constant_restitution} of nearly 1, in
order to match the quasistatic experimental regime with long-lasting
contacts (inducing a negligible amount of dissipation on short time
scales necessary for numerical stability). We also use our best
estimate of the inter-grain coefficient of friction $\mu\approx0.03$,
although with an experimental error bar of $\pm0.02$. It is likely
that friction between immersed hydrogel particles is not Coulombian.
Nevertheless, and given its very low value, we assume in the DEM that friction
is always saturated $F_{t}=\mu F_{n}$
with $F_{n}$ given by the choosen model and $F_{t}$ the magnitude of
the tangential force component (whose direction opposes sliding).
This is further justified by the fact that we discard the tangential terms
in Eq.~\ref{eq:deltakc}. For larger values of $\mu$, with non-saturated friction,
it is possible to add contributions from other contacts to $F_{t}$ 
with a similar formula~\cite{solid_mechanics}.

The following metrics are defined to quantify the simulation accuracy,
to assess the quality of each force model:

\begin{figure}[b]
\includegraphics[width=1\columnwidth]{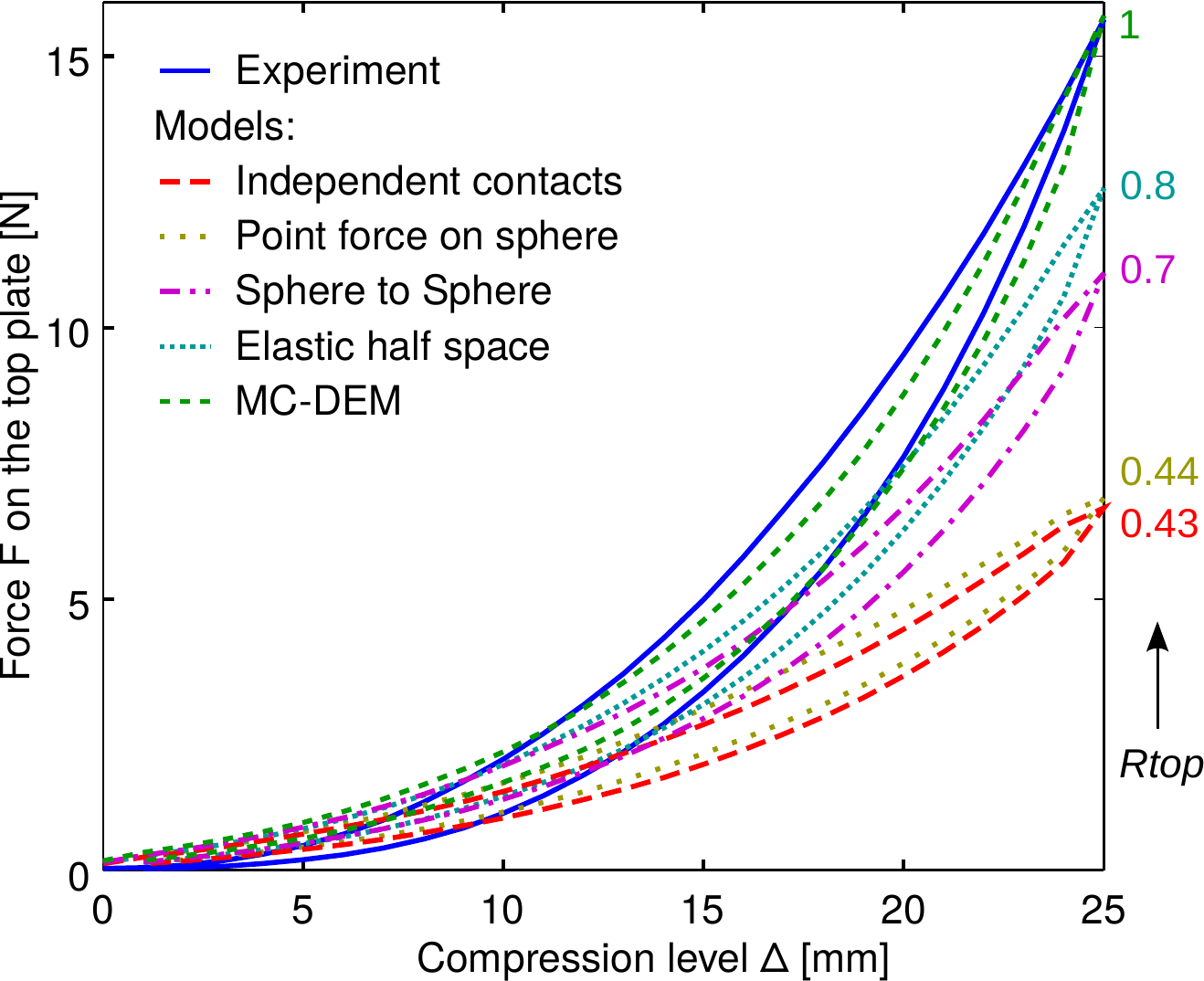}

\caption{\label{fig:Top_Forces}Force exerted on the top plate during a compression test, for
  the experimental data from~\cite{3DXP} and all the simulation models mentioned above, for $\mu=0.03$.}
\end{figure}

\textendash{}~\emph{Rtop}: The ratio between the force on the top
plate computed in the simulation and the measured one, on average, at
the maximum compression level. This value should be as close to one as
possible. Fig.~\ref{fig:Top_Forces} shows the top forces for the
experimental data and the simulation models, at all compression
levels. \emph{Rtop} is shown on the right. The
classic DEM model and the point force on sphere solutions are both off by more than
50\%. The sphere-to-sphere and linear half space models come closer,
but the geometric correction by $\gamma$ is necessary to reach a $Rtop
= 1$.

\textendash{}~\emph{errf}: The Root Mean Squared (RMS) error for
the force on the top plate at all compression levels. For simulations
with \emph{Rtop} close to 1, this measures the ability to also reproduce
correctly the hysteresis between the compression/unloading phases
visible in Fig.~\ref{fig:Top_Forces}. None of the models is able
to reproduce the experimental curve exactly, but the new MC-DEM model is
the most accurate.

\begin{figure}[b]
\includegraphics[width=1\columnwidth]{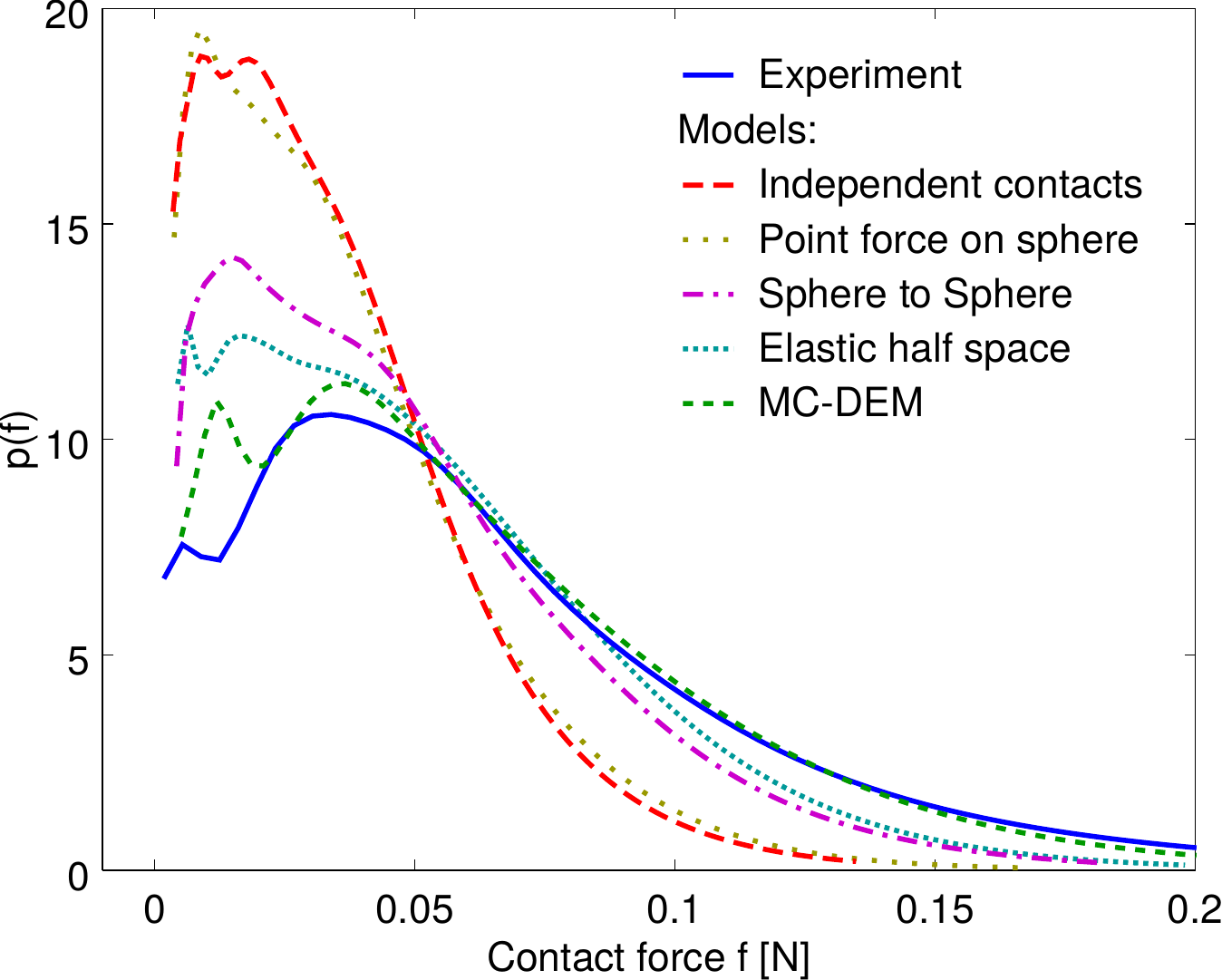}
\caption{\label{fig:fdists}Comparison of the force distributions at
  intermediate compression $\Delta=18$mm during the unloading
  phase. This representative example is chosen to show values other
  than maximal compression. The \emph{bhat} error measure in the main
  text captures all compression levels.}
\end{figure}

\textendash{}~\emph{bhat}: The above metrics only use the global measure of force
exerted on the top plate from the experimental data, but we now have access to
the full 3D forces at each contact. Due to the initial packing
rearrangement, we can only compare statistical distributions, which
the simulations must reproduce as closely as possible. One measure of
closeness for distributions is the Bhattacharyya ``distance'', defined
as $d=-log\, B$ with
$B=\int_{0}^{\infty}\sqrt{p\left(f\right)q\left(f\right)}df$, where
$p$ and $q$ are the force distributions. This definition can
consistently be averaged over all compression levels by using
$D=-\log\left\langle B\right\rangle $.  This is the measure we report
as \emph{bhat}, which is a global indicator of how closely the
distributions in the simulation match the experimental ones. $D=0$
would indicate a perfect match at all compression levels.  The value
in the least compression levels are biased by the initial packing
rearrangement, but this is the same bias for all models. There is also
an experimental lower resolution limit on weak forces which is not
present in the simulations. Fig.~\ref{fig:fdists} shows the case for
the unloading phases, at compression level $\Delta=18$mm, where the
force on the top plate is about half the maximum. As expected, the
weak forces show the most discrepancy. Nevertheless, MC-DEM
reproduces both the experimental distribution peak and the
largest force values.

\begin{figure}
\includegraphics[width=1\columnwidth]{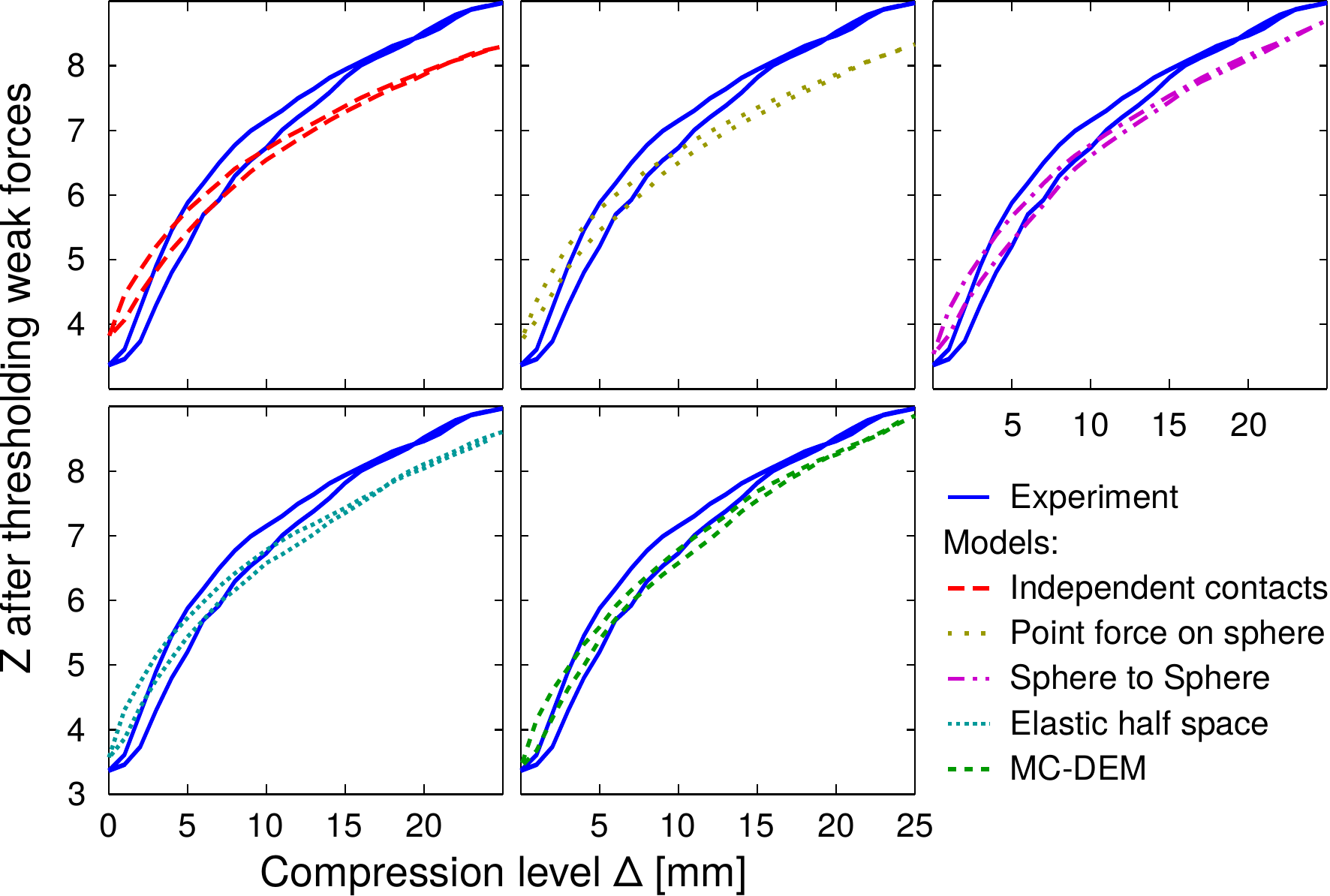}
\caption{\label{fig:thresholdedZ}Number of contacts per grain Z after
  thresholding weak forces, averaged over multiple cycles.}
\end{figure}

\textendash{}~\emph{errZ}: RMS error of the number of contacts per
grain, after thresholding weak forces $f<\tau$ in the simulations
and ignoring the corresponding contacts. The
threshold $\tau$ that produces the best match and thus lowest RMS error
for Z can be used as an estimate for the experimental lowest resolution
on the weakest forces. Here
we find $\tau=4$mN, as shown in table~\ref{tab:models_comparison}.
This is consistent with the average force in the least compression level
reported in \cite{3DXP}, $\left<f\right>=10$mN.
Fig.~\ref{fig:thresholdedZ} shows the
number of contacts per grain after thresholding for all models.
MC-DEM yields values closest to the experiment. When thresholding
is not applied, simulation models produce $6.4<Z<6.5$ in the least
compressed case: this is a side effect of the residual small
compression, as uncompressed weakly frictional grains at rest are
expected to produce values typically between 4 and 5 \cite{packing_Z}.

\begin{table}[b]
\begin{tabular}{|c|c|c|c|c|c|}
\hline 
Model & \emph{Rtop} & \emph{errf} & \emph{errZ} & \emph{bhat}$\times10^{2}$ & $\tau$ (mN)\tabularnewline
\hline 
\hline 
Independent contacts & 0.43 & 3.39 & 0.51 & 7.21 & 2.89\tabularnewline
\hline 
Point force on sphere & 0.44 & 3.23 & 0.53 & 6.51 & 2.93\tabularnewline
\hline 
Sphere to sphere & 0.70 & 1.82 & 0.34 & 2.57 & 3.35\tabularnewline
\hline 
Elastic Half Space & 0.80 & 1.26 & 0.40 & 1.90 & 3.45\tabularnewline
\hline 
MC-DEM & 1.00 & 0.41 & 0.26 & 0.98 & 4.00\tabularnewline
\hline
\hline 
\emph{Perfect fit} & \emph{1.00} & \emph{0} & \emph{0} & \emph{0} & \emph{<10} \tabularnewline
\hline
\end{tabular}
\caption{\label{tab:models_comparison}Metrics to compare DEM methods with experiment: Rtop: maximum compression force ratio at maximum compression; errf: Root Mean Squared Error of the force response at all compression levels during compression/decompression; bhat: interparticle force probability distribution function match; errZ: RMSE of the contact number per particle; $\tau$: weak force resolution.}
\end{table}

Table \ref{tab:models_comparison} summarizes all the configurations
and their performances. The traditional DEM implementation with independent contacts
clearly fails to reproduce the measurements. Implementing a linear spring model would make the match between numerics and experiments even less accurate, due to the lack of nonlinear stiffening at the contact level~\cite{DEM_recent}. The solution for a point-force on a sphere
boundary condition is not relevant in the case of multiple contacts,
and does not bring much improvement. The seminal multiple-contacts model proposed by
Gonzalez and Cuitiño~\cite{multicontacts} offers good performance, but
not as good as the half-space approximation from linear
elasticity. Since these two models are equivalent when the contact
positions are restricted to the surface of a sphere, this confirms
that even slight surface deformations are important to accurately
reproduce the experimental measures. MC-DEM, adjusting $\gamma$ so that
$Rtop=1$, yields the best results on all quantifiers. 

\section{Microscopic Variable Scaling}

\begin{figure}
\includegraphics[width=1\columnwidth]{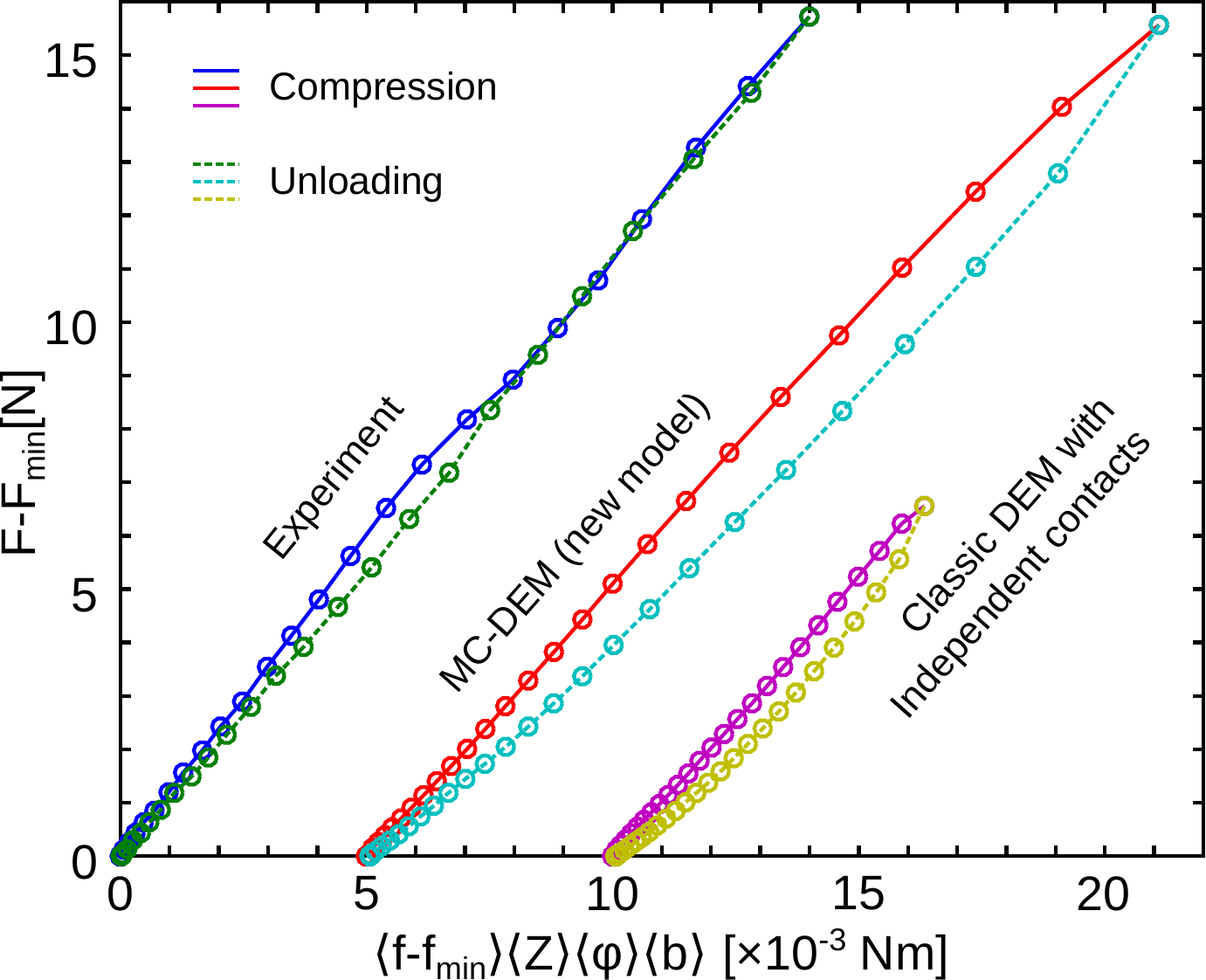}
\caption{\label{fig:scalings}Validation that the scaling law from \cite{3DXP}, derived from experimental data, also holds in the simulation. Here, $F$ is the force exerted by the grains on the top plate, $f$ is the average force at each contact, $Z$ the number of contacts per grain, $\varphi$ is the volume fraction within the packing (starting one diameter away from walls) and $b$ is the distance between grain centers. Microscopic quantities were averaged on all grains/contacts within the packing. Data was then averaged on the 15 retained compression or unloading phases with the same strain. Each marker in the diagram represents one such strain level. MC-DEM and the usual DEM data have been shifted for clarity by 5 and 10 mNm respectively.}
\end{figure}

In \cite{3DXP}, we have established a new scaling law relating the macroscopic force on the top plate to microscopic quantities, which we validated with experimental data. We summarize the scaling law here briefly: a packing stress tensor can be computed~\cite{stress_tensor} with a relation of the form $\sigma = 1/V \sum_{c \in V} \textbf{b}_c \otimes \textbf{f}_c$, with $V$ an averaging volume and $c \in V$ the contacts in that volume. For each contact $c$, $\textbf{b}_c$ is the vector between the centers of the grains and $\textbf{f}_c$ is the force vector along the contact normal. Neglecting friction and non-sphericity, $\textbf{b}_c$ and $\textbf{f}_c$ are nearly aligned; hence, the trace $tr\left(\textbf{b}_c \otimes \textbf{f}_c\right) \propto \textbf{b}_c \cdot \textbf{f}_c$. The number of terms in the sum depends on the density of contacts, which is about $\frac{1}{2} Z \phi$, with $Z$ the number of contacts per grain and $\phi$ the grain volume fraction within the packing. With density matching, we can neglect hydrostatic pressure gradients and due to sphericity, we can set the pressure on the top plate $P \propto \textbf{b}_c \cdot \textbf{f}_c$. Hence, with $F$, $f$ and averaging performed as explained in Fig.~\ref{fig:scalings}:
\begin{equation} 
F \propto P \propto \left<Z\right> \left<\phi\right> \left<\textbf{b}\right> \left<\textbf{f}\right> 
\end{equation}
The ability of the simulations to reproduce this scaling law is a strict assessement of each model quality. In Fig.~\ref{fig:scalings} we show both the experimental data, the result for MC-DEM and the result for the classic, independent contacts, DEM simulation. The new model presents an overall good agreement with the experiment data, much closer than the classic DEM scaling. Differences are a slightly lower coefficient of proportionality between the top force and the microscopic quantities, as well as more hysteresis between the compression and unloading phases. The maximum top force $F$ for both model matches the \emph{Rtop} data given in Table~\ref{tab:models_comparison}, with a very small $F_{min}$ offset. Irrespective of $F$, the scaling for the simulations presents some curvature in the unloading phase which does not show up in the experimental data (with less relative curvature for MC-DEM). One possible explanation for the discrepancy could be friction, which is assumed to follow Coulomb's law in these simulations, while this is not necessarily valid for the hydrogel particles used in the experiment.

\section{Discussion}
The new model, which we introduce here, clearly shows
that current DEM techniques, where contacts are considered to be
independent, are not very successful for simulating dense packings. We
have introduced a simple and effective way to compensate for the
geometry of the boundary conditions within the packing, in the form of
an empirical prefactor $\gamma$ for the cross-contact influences.
Ideally, $\gamma$ should be replaced by a
correct accounting for the boundary conditions within the packing, but
this is very complicated and any solution would need to be updated at
every simulation step. Working at the grain level and recoupling the
force propagation through the contact law seems the only practicable
approach for DEM. Thus, replacing the complex boundaries within the packing
with a functional dependency on statistics at grain level \cite{granocentric} would be an 
interesting option for building better models.
Future models also need to account for lower Poisson ratios $\nu<0.5$.
In the particular
case of $\nu=0$, no correlation should be introduced between
orthogonal contacts. Yet both the linear elasticity half-space
solution (\cite{solid_mechanics} and Eq.\ref{eq:deltakc}), as well as
the sphere-to-sphere contact approximation \cite{multicontacts},
predict some cross-contact displacements $\delta_{k\rightarrow c}$ in
this situation.

\section{Conclusion}
We have shown that our multiple-contact implementation in DEM captures the force dynamics of a compressed sphere packings very well, both at the macroscopic and microscopic level, including the full distribution of forces in 3D. Our approach performs significantly better than current 
DEM techniques~\cite{DEM_recent}, where contacts are considered independently, but also better than some other approaches to implementations of multiple contact modeling. We have introduced a simple
and effective way to compensate for the geometry of the boundary conditions
within the packing, in the form of an empirical prefactor $\gamma$
for the cross-contact influences. Although we compare these simulations to experiments on soft particles,
we stress that Eq.\ref{eq:deltakc} does not depend on the Young modulus
of the particles, which appears both in the force in the numerator
and in the denominator: Hard particles are less deformed when subject to
the same force, but the cross-contact displacements on the particles
have exactly the same magnitude relatively to these small deformations.
Therefore, contacts cannot be considered in isolation when simulating
dense granular packings, whatever the hardness of the grains, and
especially close to jamming, when small deformations matter. We hope
that our work will induce a chain of new MC-DEM models for these situations.

Aknowledgments and note: This work was funded in part by NASA grant
NNX10AU01G, NSF grant DMR12-06351, and ARO grant W911NF-1-11-0110.
The source code for the simulations, including the new model, is
available as free/libre software on the first author website.

\end{document}